\documentclass[prl,preprint,amsmath,amssymb,floatfix]{revtex4-1}
\usepackage{graphicx}

\usepackage{times}

\begin{document}

\title{Shot noise suppression at room temperature in atomic-scale Au junctions}

\author{P. J. Wheeler$^{1}$, J. N. Russom$^{1}$, K. Evans$^{2}$, N. S. King$^{1}$, D. Natelson$^{1, 2}$}

\affiliation{$^{1}$ Department of Physics and Astronomy, Rice University, 6100 Main St., Houston, TX 77005}
\affiliation{$^{2}$ Applied Physics Program, Rice Quantum Institute, Rice University, 6100 Main St., Houston, TX 77005}

\date{\today}

\begin{abstract} Shot noise encodes additional information not directly inferable from simple electronic transport measurements.  Previous measurements in atomic-scale metal junctions at cryogenic temperatures have shown suppression of the shot noise at particular conductance values.  This suppression demonstrates that transport in these structures proceeds via discrete quantum channels.  Using a high frequency technique, we simultaneously acquire noise data and conductance histograms in Au junctions at room temperature and ambient conditions.  We observe noise suppression at up to three conductance quanta, with possible indications of current-induced local heating and $1/f$ noise in the contact region at high biases.  These measurements demonstrate the quantum character of transport at room temperature at the atomic scale.  This technique provides an additional tool for studying dissipation and correlations in nanodevices.  \end{abstract}

%\pacs{}
\maketitle

Mechanical break junctions have proven to be an invaluable tool in
understanding the physics of electronic conduction in metals at the
atomic scale\cite{Agrait:2003}.  By bringing two metals in and out of
contact while simultaneously performing electrical measurements, it is
possible to build up a histogram of conductance values that occur when
the contact between the electrodes ranges from tunneling to the
few-atom level.  In metals with $s$-like conduction electrons,
well-defined peaks in conductance histograms near integer multiples of
the quantum of conductance ($G_{0}\equiv 2e^2/h$) frequently appear.
These are now interpreted as a signature of conductance quantization,
transport of electrons through an integer number of (ideally)
perfectly transmitting quantum channels.

The break junction approach has recently been extended to examine
ensemble-averaged conduction in single-molecule
junctions\cite{Xiao:2004,Gruter:2005,Venkataraman:2006}.  The ability
to acquire such histograms at room temperature at relatively high
rates has greatly advanced studies of molecular conduction, permitting
systematic studies of the effects of contact
functionalization\cite{Chen:2006,Park:2007}, molecular
length\cite{Venkataraman:2006}, and molecular
conformation\cite{Venkataraman:2006b}.  However, from the conductance
alone, it is generally not possible to determine whether particular
peaks in conductance histograms are the result of quantized electronic
transport through well-defined quantum channels, or whether these
peaks simply indicate particularly stable junction configurations.

The mapping of conductance into transmission coefficients for discrete
quantum electronic channels requires additional information beyond the
linear conductance.  Two approaches have been pursued at cryogenic
temperatures to infer the quantum nature of conduction in such
junctions.  Sub-gap structure in superconducting point
contacts\cite{Scheer:1998} is one means of identifying the number of
transmitting channels and their particular transmission coefficients.
Alternately, one may use shot noise to examine the same physics.  Shot
noise results from the discrete nature of the electronic charge.  In
shot noise, the mean square current noise per Hz is related to $I$,
the average DC current, as $S_{I} = F \cdot 2 e I$.  Here $e$ is the
electronic charge, and $F$ is the Fano factor.  In the classical
Poissonian limit, $F=1$.  

The Fano factor is modified in the limit of
quantum transport, when for a quantum channel $i$, electrons are
transmitted and reflected with respective probabilities, $\tau_{i}$
and $(1-\tau_{i})$.  The resulting full expression for current noise at finite
temperature in the absence of inelastic effects is\cite{Blanter:2000}
\begin{equation}
S_{I} = 4 k_{\mathrm{B}}T G_{0} \sum_{i}\tau_{i}^{2}  +  2 e V  \coth \frac{e V}{2 k_{\mathrm{B}}T}  G_{0} \sum_{i} \tau_{i}(1-\tau_{i}).
\label{eq:shotnoise}
\end{equation}
These terms include both the Nyquist-Johnson noise and shot noise contributions.  This is expected
to be valid at frequencies much smaller than $k_{\mathrm{B}}T/\hbar$
and $eV/\hbar$, and voltages smaller than the characteristic energy
scales over which the $\tau_{i}$ vary.  As is clear from
Eq.~(\ref{eq:shotnoise}), in the limit of a small number of channels,
one expects strong suppression of noise when channels are fully
transmitting ($\tau_{i} \rightarrow 1$).  Shot noise is further
suppressed in macroscopic conductors due to inelastic electron-phonon
scattering, which effectively ``smears" electrons out across multiple
channels\cite{Blanter:2000}.

Following groundbreaking work in semiconductor quantum point
contacts\cite{Reznikov:1995,Kumar:1996}, low temperature (4.2~K)
measurements of shot noise in Au break junctions\cite{vandenBrom:1999}
have found shot noise suppression near integer conductance values,
demonstrating that the conductance quantization observed in those
experiments does result from fully transmitting channels.  More recent
measurements\cite{Djukic:2006} have used shot noise to infer the
transmission coefficients in the dominant channels in transport
through single deuterium molecules.  In these experiments noise is
measured using a low bandwidth approach involving cross-correlation of
voltage amplifiers, necessitating very stable junctions for the
comparatively slow acquisition.

Here we present measurements of noise in Au breakjunctions
acquired at room temperature using a high frequency approach
coincident with conductance measurements.  We observe peaks in
conductance histograms at the roughly 1, 2, and 3$G_{0}$, as expected
for this material.  The averaged noise shows suppression at precisely
the same conductance values as the peaks in the conductance
histograms.  This demonstrates that the quantum mechanical suppression
of noise due to fully transmitting contacts survives even at room
temperature, where inelastic processes are much stronger than in
previous low temperature experiments.  We discuss the measurement
technique and its limitations, including indications of possible $1/f$
contributions and Joule heating-induced Johnson noise.  These
experiments raise the possibility of performing room temperature shot
noise measurements in single molecules.

%Mechanical break junction consisting of a notched Au wire mounted
%on metal shim-stock.  Actuation achieved via zo actuator.
%Automated data collection.

\begin{figure}[h!]
\includegraphics[clip, width=8cm]{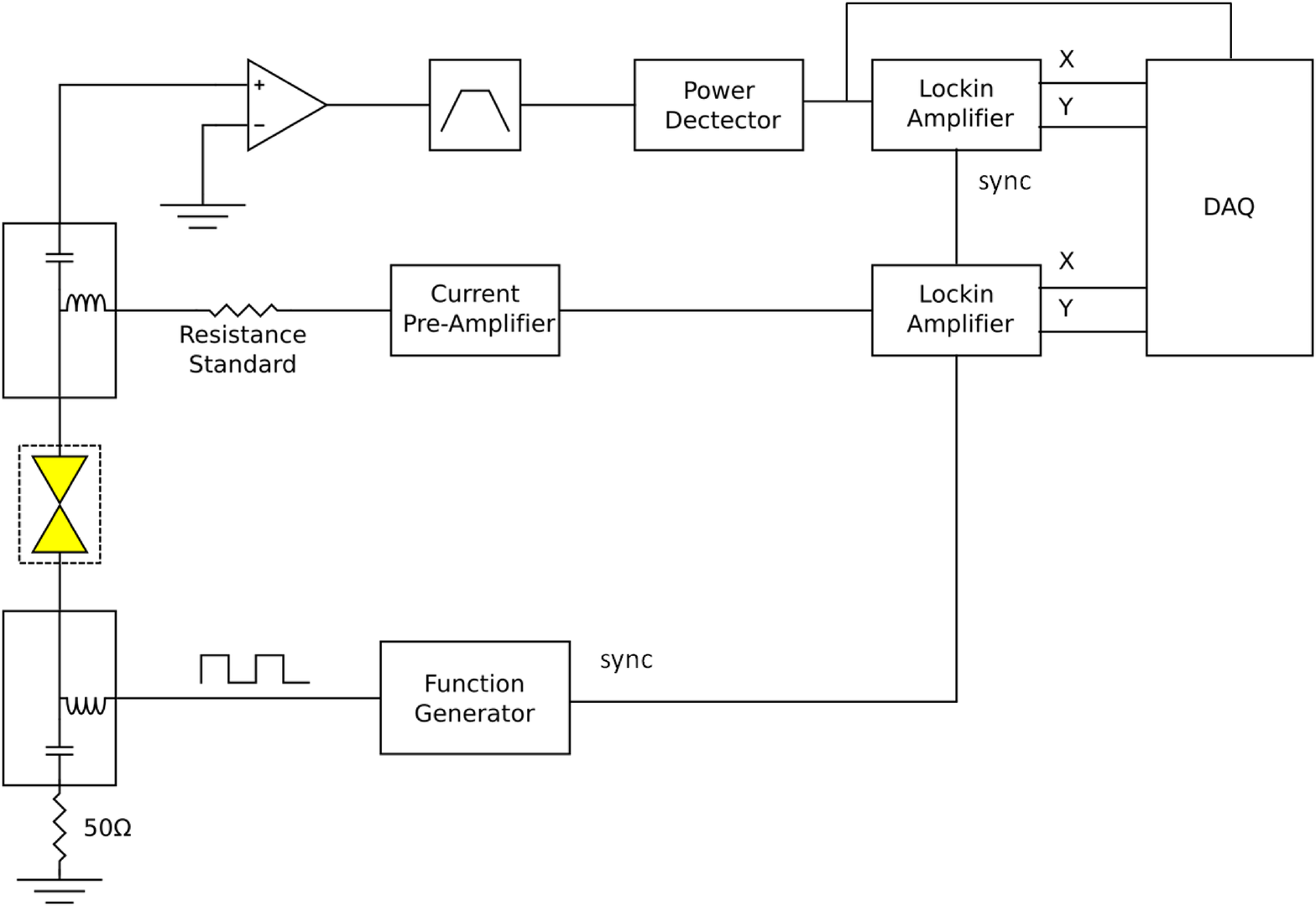}
\caption{Schematic of measurement apparatus.  The dashed line indicates
components grouped together within a shielded enclosure.}
\label{fig1}
\end{figure}

Figure~\ref{fig1} shows a schematic of the measurement setup.  A
mechanical break junction is formed using a notched gold wire mounted
on metal shim-stock.  A computer-controlled piezo motor is used to
flex the substrate, bringing the ends of the gold wire repeatedly in
and out of contact.  These measurements are performed under ambient
conditions at room temperature.  First we consider the ``DC'' portion
of the measurement circuit.  To determine the conductance, a function
generator sources an offset square wave of desired amplitude so that
the applied voltage oscillates between zero and a set value,
$V_{\mathrm{DC}}$.  This voltage is passed into the junction via the
``DC'' input of a bias-tee.  The frequency of the square wave is
typically below 10~kHz, a bound set by the properties of the bias-tee
used to deliver the voltage to the junction.  The other side of the
junction is connected through a current-limiting resistance standard
(General Radio 1433-G Decade Resistor) to a Keithley 428-PROG current
amplifier.  The resistance standard is used to avoid overloading the
current amplifier input stage when the junction is in a fully shorted
high conductance configuration.  The output of the current amplifier
is measured using a lock-in amplifier synchronized to the input square
wave, with a typical lock-in output time constant of $\sim$1~ms.

The shot noise measurement relies on the high frequency response of
the circuit, which is very different from that at DC.  The bias-tees
play a crucial role here.  The side of the junction that is biased at
DC via the function generator is terminated at 50~$\Omega$ at high
frequencies.  Similarly, at high frequencies the other bias-tee
isolates the other side of the junction from the resistance standard
and the current preamplifier.  Instead, that side of the junction is
coupled via an extremely short (few cm) piece of 50~$\Omega$ coaxial
cable directly to the input of the RF amplifier chain with a nominal
gain of $\sim$70~dB, followed by filters that define a nominal
bandwidth extending from 250 to 520~MHz.  The RF signal is then passed
to a power meter with an output proportional to the logarithm of the
incident RF power.  The output of this RF power meter is detected
using a second lock-in amplifier also synchronized to the input square
wave, with a lock-in output time constant of $\sim$1~ms.

Note that one cannot think of the RF circuit in quasi-DC terms, with
the resistance of the junction and the stray capacitance forming an RC
divider that suppresses all high frequency response.  Instead one must
consider the junction as a source of mean square current fluctuations,
$I_{s}^{2}$, in parallel with some characteristic impedance, $Z_{s}$,
determined by resistive and reactive components.  As explained in the
Supporting Information, one end of the junction is terminated at
50~$\Omega$, and the terminated junction acts as an equivalent noise
source with an effective impedance $Z_{g}=Z_{s}+Z_{0}$.  This
``generator'' is then connected to a very short piece of $Z_{0}\equiv
50~\Omega$ transmission line to a nominally 50~$\Omega$ amplifier
chain.  While \textit{maximal} power transfer would require
$Z_{g}=Z_{0}$, power is still delivered to the amplifier chain even
when there is an impedance mismatch.  If $Z_{g}>>Z_{0}$, the power
transferred to the amplifier chain is approximately $I_{s}^{2}Z_{0}$.
At any given frequency within the nominal bandwidth of the RF system,
one can measure the effects of the impedance mismatch by measuring the
reflection properties of the junction.  Ideally one wants to do this
simultaneously with data acquisition, across the whole bandwidth, but
that is not practical, and that data is acquired separately.

\begin{figure}[h!]
\includegraphics[clip, width=12cm]{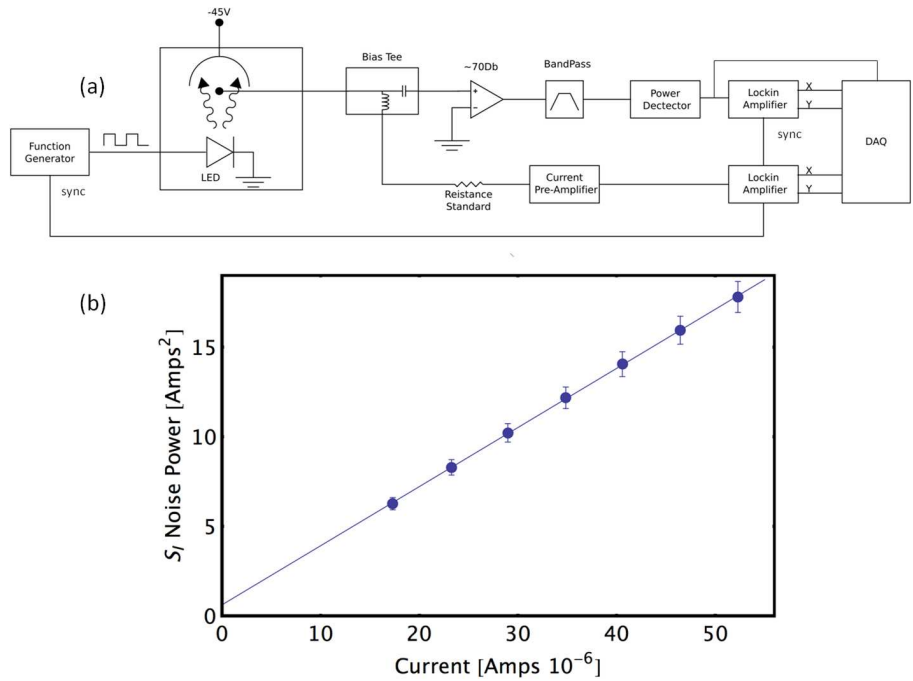}
\caption{(a) Schematic of measurement set-up for measuring the shot noise of a vacuum photodiode.  Supporting information gives further description of how the effective gain-bandwidth product is calibrated.  (b) Shot noise [A$^{2}$/Hz] as a function of (square wave) current through the photodiode (tuned by different LED illumination intensities).  The small offset indicates a small noise background, while the slope is consistent with the expected $2 e I$ to within 3\%.}
\label{fig2}
\end{figure}

We demonstrate that this approach is capable of measuring shot noise
in a system with a high DC impedance, provided the RF properties are
calibrated.  As shown in Fig.~\ref{fig2}, we have use the same
amplifier system to measure the shot noise produced by a vacuum
photodiode.  The photodiode cathode is biased through a bias-tee to
-45~V using a DC power supply, while the RF port of the cathode
bias-tee is terminated at 50~$\Omega$. The anode is connected through
the DC terminal of a bias-tee to the current amplifier and subsequent
lock-in.  The RF port of the anode bias-tee is then connected to the
amplifier chain.  A light-emitting diode (LED) is driven by the square
wave output of a function generator to illuminate the photodiode.  The
current amplifier measures the (quasi-DC when LED is on) photocurrent.
The RF amplifier chain amplifies the noise power, which is then
measured by the power meter.  Lock-in detection (synced to the LED
illumination) is then used to find the noise power produced by the
photocurrent.

Depending on illumination level, the DC differential resistance of the
vacuum photodiode around the operating point ranges from 28 to
205~M$\Omega$.  However, the RF properties of the photodiode rather
than this DC resistance determine the power transfer of the shot noise
to the amplifier chain.  As described in the Supporting Information, a
reflectometry measurement of the cathode-terminated photodiode from
the anode side is sufficient to characterize this power transfer, via
a frequency-dependent reflection coefficient, $|\Gamma(\omega)|^{2}$.

Conversion of the measured noise power into units of current noise
per unit bandwidth requires knowledge of both $|\Gamma|^{2}$ and the
gain-bandwidth product, $g$, of the amplifier chain.  An RF source is
cycled through 1000 evenly spaced frequencies, $f$, from 1~MHz to
1~GHz, feeding an input power, $P_{\mathrm{in}}(f) = -80$~dbm, into
the amplifier chain.  The output of the amplifier chain,
$P_{\mathrm{fg}}(f)$ is recorded at each frequency, as is the
background, $P_{\mathrm{bg}}(f) \equiv$ output with
$P_{\mathrm{in}}(f)$ set to zero.  The gain-bandwidth product is 
then computed by numerical integration:
\begin{equation}
g = \int \frac{P_{\mathrm{fg}}(f)-P_{\mathrm{bg}}(f)}{P_{\mathrm{in}}(f)}df
\end{equation}
A separate measurement of $|\Gamma(\omega)|^2$ of the photodiode at
each frequency (see Supporting Information) allows us to correct $g$
for the division of power between the photodiode and the amplifier
chain.  The measured noise power is then divided by this corrected
gain-bandwidth product to infer the power noise per Hz delivered to
the amplifier chain.  Finally this is then divided by the input
impedance of the first stage amplifier, 50~$\Omega$, to obtain the current
noise per Hz produced by the photodiode.  As shown, this noise scales
linearly with the photocurrent, with a slope of $3.29 \pm 0.02 \times
10^{-19}$ Amps/Hz.  This compares well with expectations, deviating
from the expected $2e$ by less than 3\%.  This demonstrates that the
RF approach can successfully measure shot noise, even though the DC
resistance of the photodiode is much greater than 50~$\Omega$.

In the break junction case, lock-in detection of the RF power is
crucial, since at 300~K the Nyquist-Johnson equilibrium noise is
significant.  Neglecting quantum effects, the Nyquist-Johnson current
noise of a 1~$G_{0}$ conductor at 300~K is $1.28 \times
10^{-24}$~A$^{2}$/Hz, while (for $F=1$) the current shot noise at a DC
current of 1~$\mu$A would be $3.2 \times 10^{-25}$~A$^{2}$/Hz.
Without lock-in detection, the shot noise would barely be detectable
above the thermal noise at a DC bias voltage exceeding 100~mV.  Even
with phase-sensitive detection there is some background RF noise power
even at $V_{\mathrm{DC}} = 0$.  This background is strongly reduced
when great care is taken on shielding extraneous interference.  While
the logarithmic power meter allows noise measurements over a large
dynamic range, it is particularly sensitive to background noise,
compared with a linear power detector.  As we discuss further below,
lock-in detection of the noise power avoids contributions from the
\textit{equilibrium} thermal noise; however, heating and resultant
temperature changes due to the applied DC bias would be detected in
this approach.  Thus, the measured noise power is likely a combination
of shot noise and bias-induced-heating changes in the Nyquist-Johnson noise.
An additional possible contribution due to $1/f$ ``flicker''
noise will be discussed below.

The reflection coefficient measurement is considerably more
challenging for the break junction case.  In principle one would like
to measure $|\Gamma(\omega)|^2$ at frequencies across the RF bandwidth
on-the-fly.  In practice, one can acquire histograms of
$|\Gamma(\omega)|^2$ as a function of $G$ at a series of discrete
frequencies.  Doing this densely across the whole frequency range of
interest is laborious and difficult.  Measurements indicate (see SI)
that at individual frequencies, $|\Gamma(\omega)|^2$ is essentially
independent of $G$ for conductances between $\sim 0.5$ and 5~$G_{0}$.
In the break junction data presented here we choose to plot the noise
in ``arbitrary units'', accepting that our incomplete knowledge of
$\Gamma(\omega)$ constitutes a systematic error (independent of $G$
from 0.5 to 5 $G_{0}$) in the overall magnitude of the noise.

Conductance and noise data are recorded simultaneously as the junction
is brought into and out of contact, using a DAQ system to sample the
lock-in outputs at $\sim$1k samples/s.  Histograms of the measured
conductance values are compiled in real time during junction breaking
and formation.  Simultaneously, running averages of the noise power
are computed for each conductance bin.  The data analysis is entirely
automated, with no post-selection of ``nice'' traces.  In general,
conductance histograms were cleaner, with more observable conductance
quanta, during junction breaking rather than formation.

\begin{figure}[h!]
\includegraphics[clip, width=8cm]{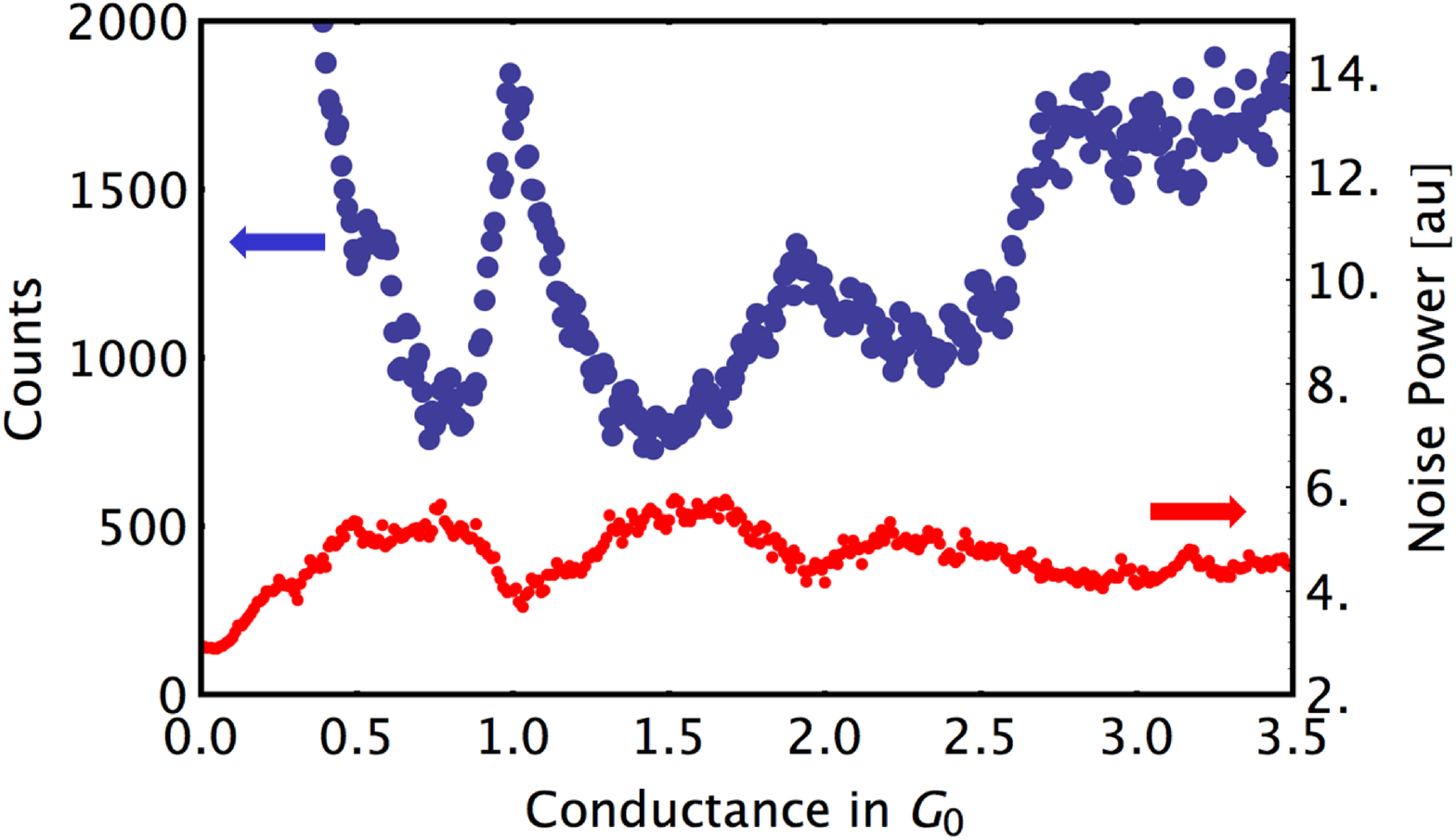}
\caption{Simultaneously acquired conductance histogram (upper curve,
 left axis) and averaged shot noise (lower curve, right axis).  These
 data were taken at room temperature with $V_{\mathrm{DC}} = 100$~mV,
 and square wave frequency $\sim$1kHz.}
\label{fig3}
\end{figure}

Figure~\ref{fig3} shows an example data set, acquired at room
temperature over the course of several hours.  The conductance
histogram clearly shows peaks at approximately 1, 2, and 3~$G_{0}$, as
expected from previous work on Au junctions\cite{Venkataraman:2006}.
Deviations from perfect quantization at integer multiples are seen in
some samples, but are always quantitatively consistent with previous
measurements on Au junctions at room temperature in the presence of
work hardening\cite{Yanson:2005}.  The averaged noise power
distribution shows significant dips in power centered at precisely the
conductance values associated with peaks in the histogram.  This noise
suppression is very similar to that seen by van den Brom \textit{et
  al}. at liquid helium temperatures\cite{vandenBrom:1999}.  Note that
the suppression is not expected to reach all the way to $F=0$, since
statistically junctions with $G = G_{0}$ may be formed in many ways,
with many possible combinations of $\tau_{i}$ for different channels.

%Still, the noise suppression shows that the peaks in the conductance histogram do %represent a preponderance of junction configurations with integer numbers of %channels with $\tau_{i} \rightarrow 1$.

\begin{figure}
\includegraphics[clip, width=12cm]{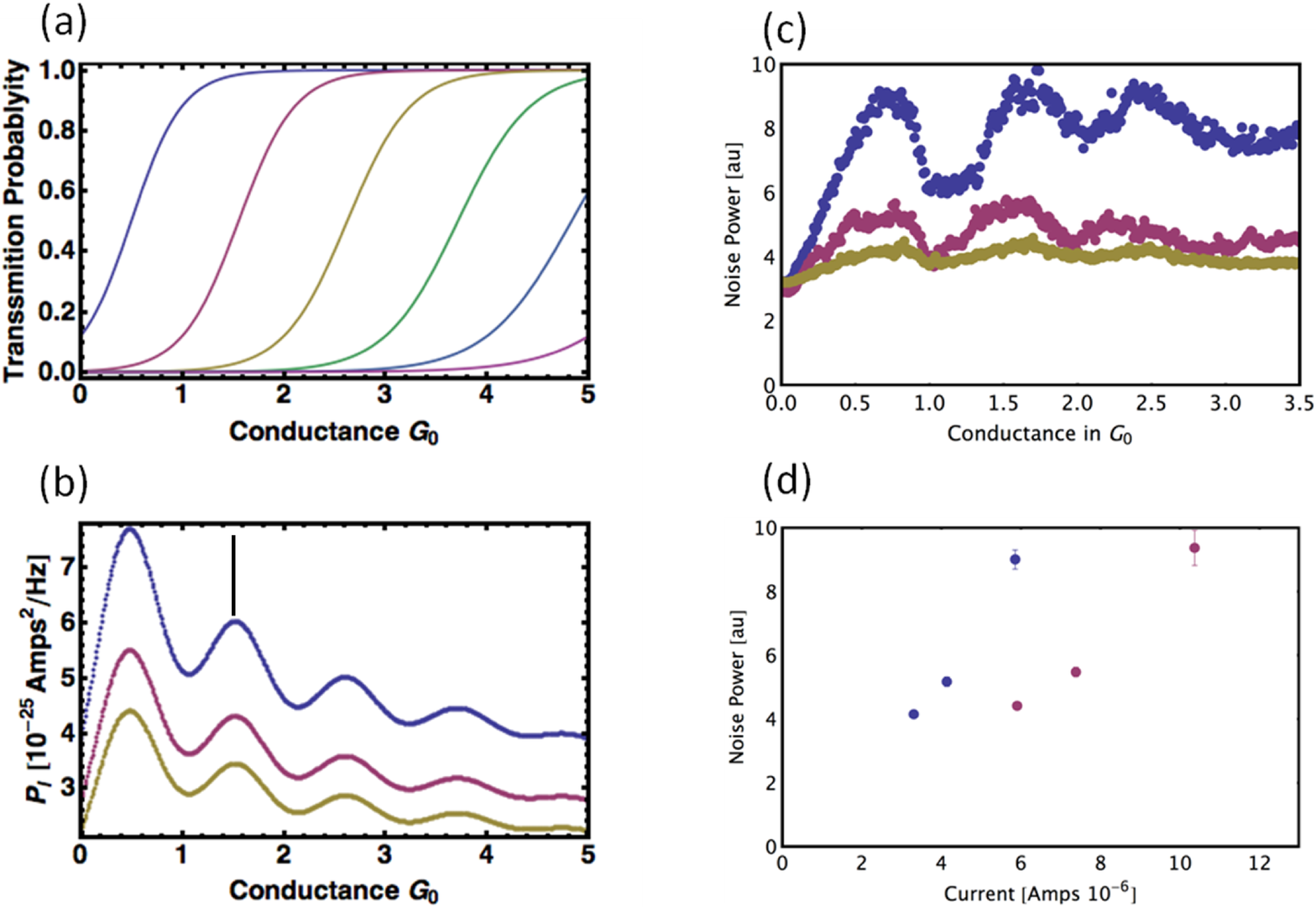}
\caption{(a) One possible choice of relevant transmission coefficients as a function of junction conductance.  This is similar to the approach taken by van den Brom and van Ruitenbeek\protect{\cite{vandenBrom:1999}}. 
(b) Predicted current noise with no Nyquist-Johnson component, for (bottom to top) $V_{\mathrm{DC}} = 80$~mV, 100~mV, and 140~mV, calculated from Eq.~(\ref{eq:shotnoise}) using the transmission coefficient distribution shown in (a) and recalling that $V$ across the junction varies with $G$ due to the series resistor.  The vertical line shows the equivalent change in current noise expected at that 
conductance for a temperature change of 20~K.
(c) Measured noise power as a function of conductance for (bottom to top) $V_{\mathrm{DC}} =$~80~mV, 100~mV, and 140~mV on a particular break junction, with no background subtraction. 
(d) Measured noise power as a function of DC bias current at two different conductances.  Dependence on the DC current is slightly superlinear.
Possible explanations for this are discussed in the text. }
\label{fig4}
\end{figure}

As described by van den Brom and van Ruitenbeek\cite{vandenBrom:1999},
one can construct an explicit model of the expected shot noise by
assuming a form for the contributions of different conductance
channels as a function of conductance.  This is likely to be
reasonable only in the small number of channels limit.
Figure~\ref{fig4}a shows such a model, while Fig.~\ref{fig4}b plots
the corresponding expected shot noise, based on
Eq.~(\ref{eq:shotnoise}), for different magnitudes of DC (square-wave)
bias.  The calculations assume a temperature of 300~K, and take into
account that the actual voltage dropped across the junction is $V =
V_{\mathrm{DC}} (1/G)/(1/G + R_{\mathrm{ser}})$, where
$R_{\mathrm{ser}}$ is the resistance standard, approximately
6~k$\Omega$ for the measurements in Figs.~\ref{fig3} and \ref{fig4}c.

There is considerable interest in comparing data sets taken at
different junction biases.  Higher biases enable possible inelastic
processes such as optical phonon scattering, and there are theoretical
predictions\cite{Chen:2003,Chen:2005} (particularly in molecular
junctions\cite{Mitra:2004,Koch:2005}) that such processes can modify
the Fano factor away from the prediction of Eq.~(\ref{eq:shotnoise}).
Magnetic processes can also lead to unconventional Fano factor
values\cite{Sela:2006}.  The calculation shown in Fig.~\ref{fig4}b
assumes no such modifications.

Figure \ref{fig4}c shows averaged measured noise power as a function
of conductance for the same $V_{\mathrm{DC}}$ magnitudes as in the
model of Fig.~\ref{fig4}b.  No background subtraction has been
performed, and as above no correction has been made to the noise data
to account for the dependence of $|\Gamma(\omega)|^2$ of the junction
on the conductance.

%The background noise for each run measured at $G = 0$ has been
%subtracted from %each bias voltage data set.  When $G = 0$, there can
%be neither shot noise nor %Johnson noise, implying that any residual
%noise detected is pickup from extraneous %sources.
%% 

Comparison between the model and the data reveals several features of
interest.  The overall trend at high conductances toward lower noise
with increasing $G$ is expected due to the resistance standard: When
$G >> R_{\mathrm{ser}}^{-1}$, the DC current approaches
$V_{\mathrm{DC}}/R_{\mathrm{ser}}$, and the DC voltage, $V$, relevant
for Eq.~(\ref{eq:shotnoise}) becomes small because most of the applied
DC voltage is dropped across the resistance standard.  The deviation
of the data from the model at low conductances ($G < \sim 0.5 G_{0}$)
is not surprising, given the sensitivity of that data to the
reflection coefficient resulting from the impedance mismatch between
the junction and the amplifier chain.

One must be concerned about contributions to the data due to $1/f$
noise.  In electronic conduction this noise results from temporal
fluctuations of the resistance, with a broad distribution of
characteristic timescales\cite{Dutta:1981,Weissman:1988}.  This noise
has been shown to exist in metal contacts approaching the atomic
scale\cite{Wu:2008}.  Static conductance fluctuations as a function of
DC bias voltage across nanojunctions are known to be suppressed near
quantized conductance values\cite{Ludoph:1999}.  It is conceivable
that the time-dependent conductance fluctuations that give rise to
$1/f$ noise could be similarly affected by the decreased
backscattering and increased junction stability near quantized
conductance values.  To test for this physics, we consider the
dependence of the measured noise on the DC current through the
junction.  At a given $G$, conductance fluctuation noise is expected
to be \textit{quadratic} in the DC current, in contrast to shot noise.
We do not see strong signatures of such a dependence, as shown in
Fig.~\ref{fig4}d, finding that the measured noise is roughly linear in
DC current, extrapolating toward a finite background value at zero
current.  There is some nonlinearity, however, meaning that some
contribution from $1/f$ noise cannot be completely ruled out.

The presence of noise suppression at room temperature demonstrates
explicitly the quantum character of transport in these atomic-scale
devices.  Inelastic processes such as electron-phonon scattering can
remove energy from the ``hot'' electron system and redistribute
electrons between the different quantum channels.  This is what leads
to the suppression of shot noise in macroscopic conductors at room
temperature.  These data suggest that such inelastic processes operate
on length scales longer than the single-nanometer junction size, even
at 300~K and biases in the tens of mV range.

An additional contribution to the superlinearity seen in
Fig.~\ref{fig4}d that must be considered is bias-driven local
electronic heating of the junction.  Such heating would be synchronous
with the bias current, and thus would be detected in the lock-in
technique.  Bias-driven ionic heating has been inferred in such
junctions previously by studying the bias-dependence of the junction
breaking process\cite{Tsutsui:2007}.  It has also been noted that
effective temperature changes in robust junctions can reach several
hundred Kelvin at biases of a few hundred mV\cite{Ward:2008}.  An
effective increase in the electronic temperature (which is relevant to
Eq.~(\ref{eq:shotnoise}) would not have to be very large to be
detectable.  The plausibility of this explanation is reinforced by the
vertical line shown in Fig.~\ref{fig4}b, indicating the change in
Nyquist-Johnson current noise expected for an electronic temperature
change at that conductance of 20~K.

Such electron heating has been considered theoretically in some
detail\cite{Chen:2003,Chen:2005,DAgosta:2006}.  In fact, the authors
of Ref.~\cite{DAgosta:2006} explicitly suggest using measurements of
the noise as a means of detecting local, nonequilibrium heating in
atomic-scale contacts.  Detailed modeling of the local junction
temperature is complicated by the fact that local Joule heating and
effective thermal path are explicit and implicit functions,
respectively, of $G$.  Such modeling is beyond the scope of the
present paper and will be examined carefully in a later publication.

%An alternative candidate explanation for these voltage-dependent
%trends would be the changes in the effective Fano factor due to
%inelastic processes mentioned above.  However, this seems quite
%unlikely in metallic junctions, given previous measurements at
%cryogenic temperatures and the persistence of the elevated noise over
%a broad range of conductances.

This technique raises the possibility of performing rapid assays of
shot noise through molecules, given the great progress that has been
made in recent molecular break junction conductance measurements.
As mentioned above, both electron-vibrational effects and magnetic
processes are predicted to modify Fano factors away from $F=1$.
However, small molecule conductances tend to be on the order of
$10^{-5}-10^{-2}~G_{0}$.  Such low conductance junctions have,
necessarily, very poor impedance matching to the usual 50~$\Omega$ RF
electronics used in the noise measurements.  Moreover, there is great
interest in examining noise in such systems at much lower bias
currents.  Improved coupling of junction noise power to the amplifiers
and reduced backgrounds would be a necessity.  Impedance matching
networks or tank circuits may provide a means of adapting this
approach to the molecular regime, though not without a likely
reduction in bandwidth.

We have used high frequency methods to observe shot noise suppression
in atomic-scale Au contacts at room temperature, demonstrating clearly
the quantum character of conduction in these nanodevices.  High
frequency methods allow the rapid acquisition of noise data
simultaneously with statistical information about conduction in
ensembles of junctions.  A slightly nonlinear dependence of the
measured noise on DC bias current raises the possibility that local heating and $1/f$ noise may need to be considered in these structures.
While impedance matching for low conductance junctions is a challenge,
the prospect of adapting this approach to study noise in molecular
junctions is appealing.

%[Insert concluding paragraph.]

DN and PJW acknowledge support from Robert A. Welch Foundation
Grant C-1636 and NSF grant DMR-0855607.

\clearpage

\clearpage

\clearpage

\clearpage

\clearpage

\textbf{Supporting Information}
\\

\renewcommand{\thefigure}{S\arabic{figure}}

This document contains supporting information regarding RF
measurements of shot noise reported in the above manuscript.

\section{RF equivalent circuits}

\begin{figure}[h!]
\includegraphics[clip, width=12cm]{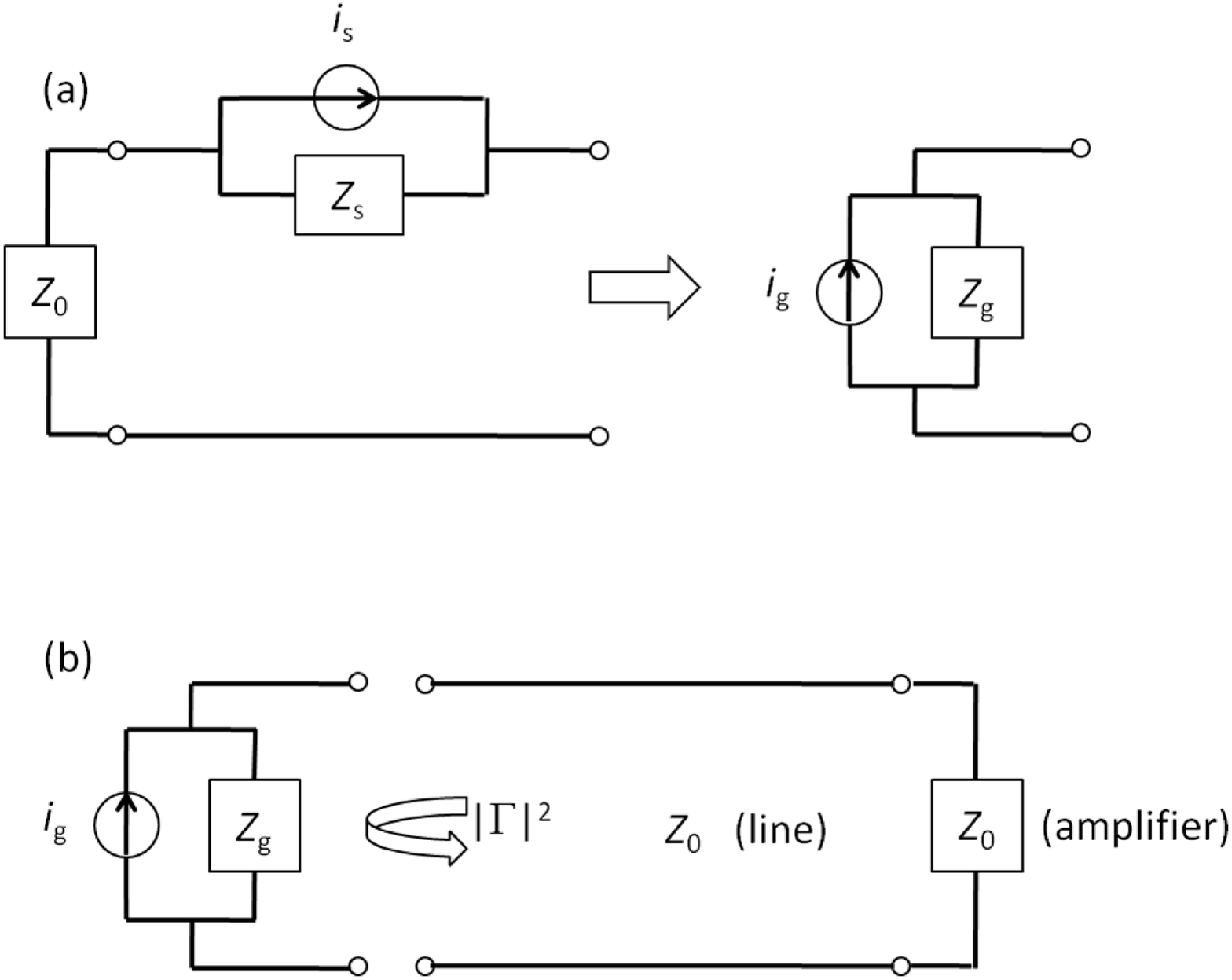}
\caption{(a) Equivalent circuits for noise current sources.  The sample itself has some short-circuit rms current noise, $i_{s}$, and effective impedance, $Z_{s}$.  Terminating one end of the sample with $Z_{0}$ results in a single-port equivalent circuit with rms current noise $i_{g}$ and impedance $Z_{g}$, as explained in the supporting text.  (b) Schematic of the reflection measurement that is relevant to determining the actual power transferred from the sample to the amplifier chain.}
\label{sfig1}
\end{figure}

As shown in Fig.~\ref{sfig1}a, one should think of a noise source
(either a break junction or a vacuum photodiode, in the cases
considered here) as an ideal current source of mean square current
fluctuations $i_{s}^{2}$ in parallel with an impedance $Z_{s}$ that
depends on frequency.  The actual noise source is a two-port device
at RF, terminated at one end by $Z_{0} = 50~\Omega$.  The 
equivalent circuit parameters for the resulting single-port
``generator'' are then 
$$i_{g} = i_{s} \frac{Z_{s}}{Z_{s}+Z_{0}},$$
$$Z_{g}=Z_{s}+Z_{0}.$$

One then considers attaching this generator to a transmission line
(characteristic impedance $Z_{0}$) and the RF amplifier chain (also
$Z_{0}$).  The current delivered to the $Z_{0}$ load is $i_{0} = i_{g}
Z_{g}/(Z_{g}+Z_{0})$, and the power transferred to the $Z_{0}$ load is
then 
$$|i_{0}|^{2}Z_{0} = Z_{0}i_{s}^{2} |\frac{Z_{s}}{2Z_{0}+Z_{s}}|^{2}.$$

If one considered a reflectance measurement looking from a 50~$\Omega$
line into the generator (Fig.~\ref{sfig1}b), one would find
$$\Gamma = \frac{Z_{g}-Z_{0}}{Z_{g}+Z_{0}} =
\frac{Z_{s}}{2Z_{0}+Z_{s}}.$$
Thus, the power transferred to the $Z_{0}$ load is
$i_{s}^{2}Z_{0}|\Gamma|^{2}$.
In the limit that $Z_{s}>>Z_{0}$, $\Gamma \rightarrow 1$. 

\section{Photodiode measurements}

\begin{figure}[h!]
\includegraphics[clip, width=12cm]{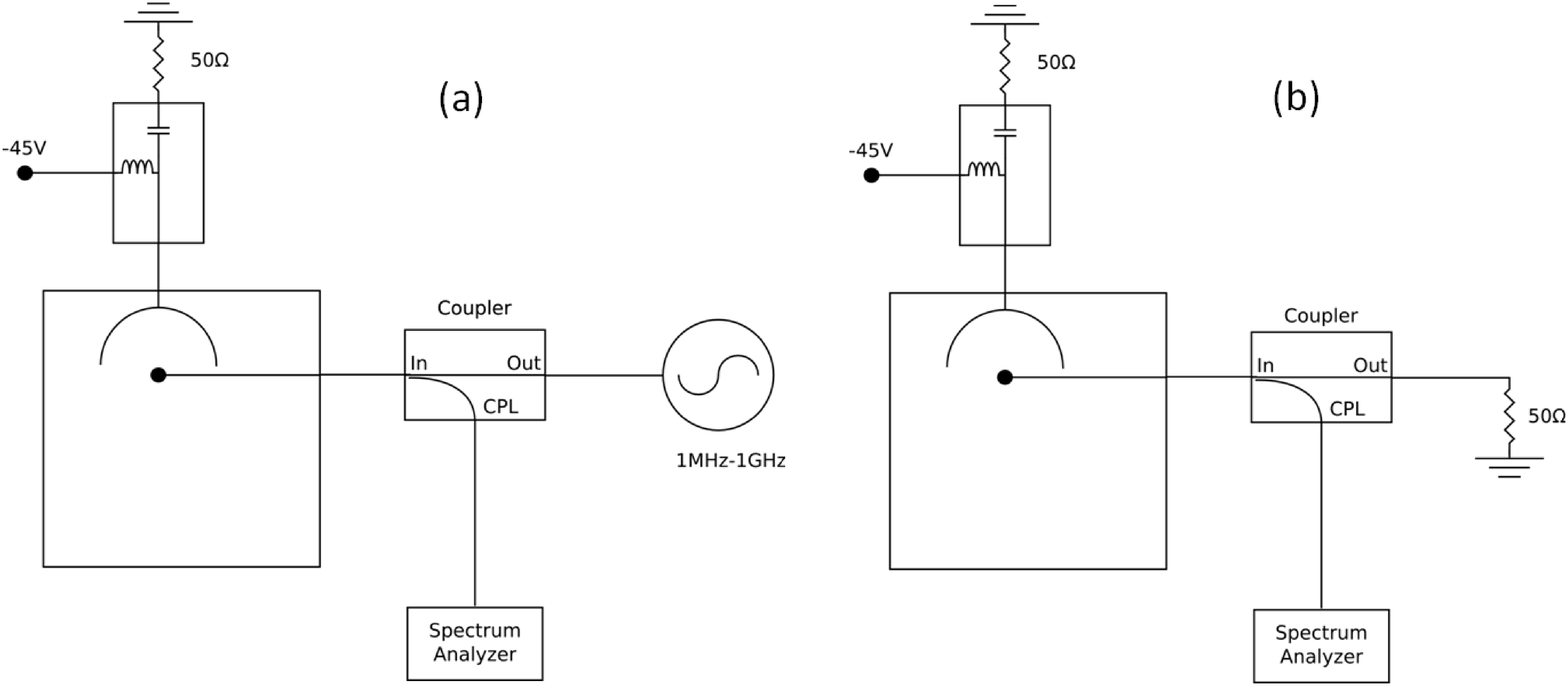}
\caption{Schematics of circuits used to measure the RF reflection coefficient of the vacuum photodiode over the bandwidth of interest.  The directional coupler has a directional throughput of -20~dB.  The left circuit measures the reflected power at fixed frequency, and the spectrum analyzer's RF source is swept through 1000 discrete frequencies.  The right circuit measures the background noise power of the system.}
\label{sfig2}
\end{figure}

Fig.~\ref{sfig2} shows the circuit configurations used to measure
the reflection coefficient of the terminated vacuum photodiode.

\begin{figure}[h!]
\includegraphics[clip, width=12cm]{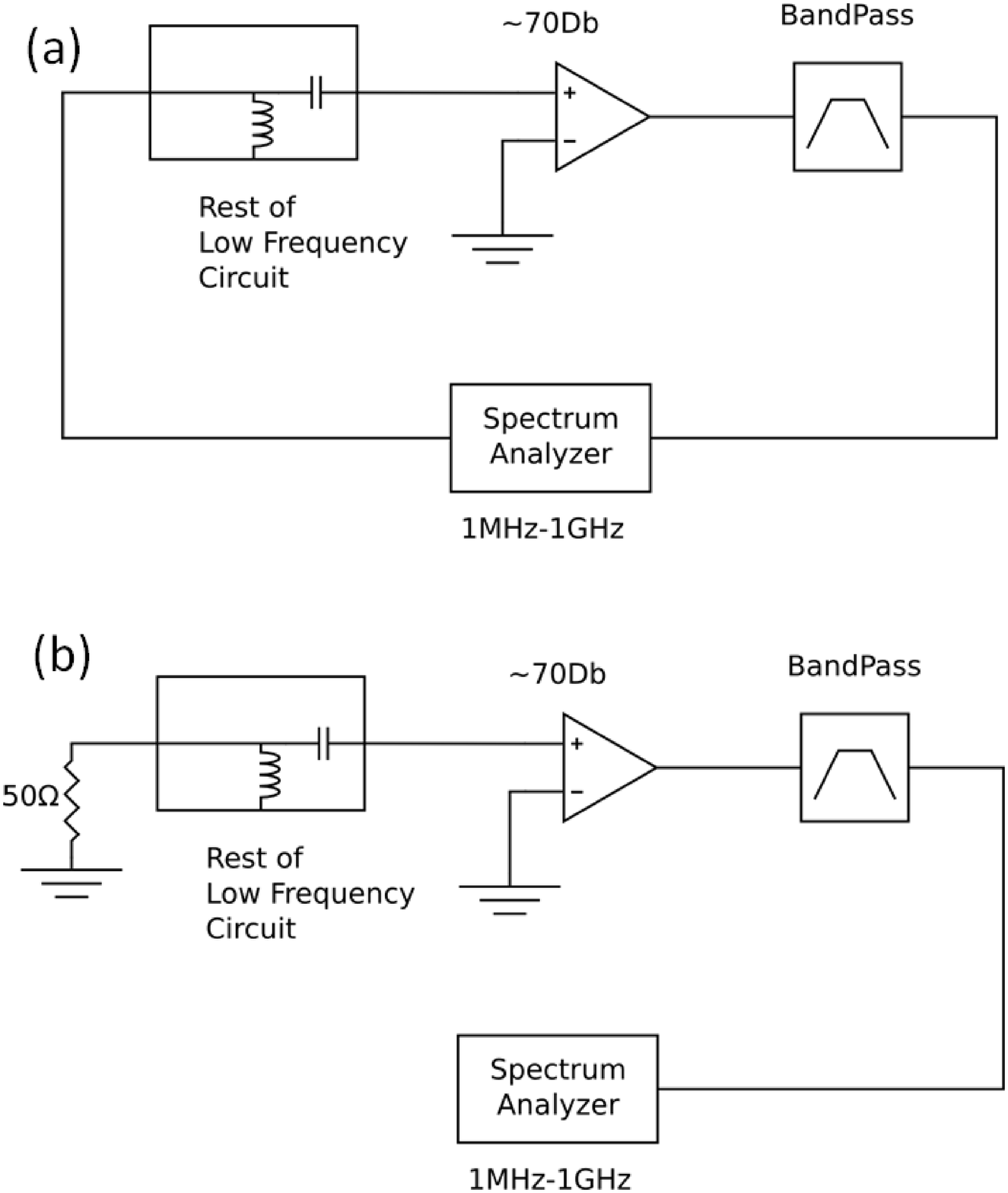}
\caption{(a) Schematic diagram showing setup for measuring the gain-bandwidth product of the amplifier chain used in the experiments.  The ``rest of low frequency circuit'' branch includes the resistance standard, current amplifier, and lock-in amplifier shown in Fig.~1 of the full manuscript. (b) Schematic diagram showing circuit for background noise measurement in gain-bandwidth product configuration.}
\label{sfig3}
\end{figure}

Fig.~\ref{sfig3} shows the circuit configurations used to measure the
gain-bandwidth product of the nominally 50~$\Omega$ amplifier chain.
Gain was measured by comparing output power referenced to a known
input power for 1000 discrete frequencies across the bandwidth 
of interest.

\begin{figure}[h!]
\includegraphics[clip, width=12cm]{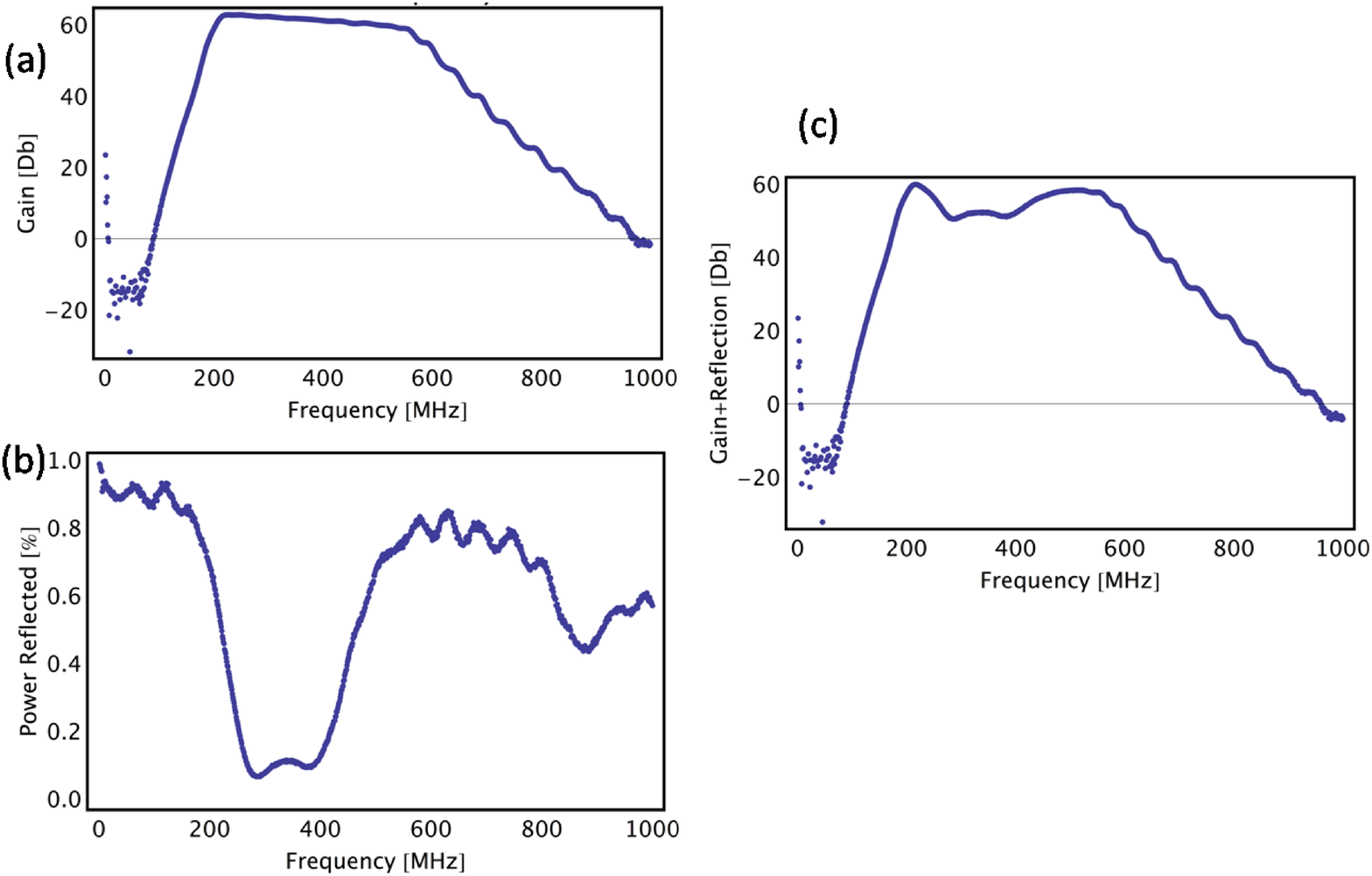}
\caption{(a) Measured gain vs. frequency obtained using the circuit in Fig.~\ref{sfig3}.  (b) Fraction of power reflected by the photodiode as a function of frequency, obtained using the circuit in Fig.~\ref{sfig2}.  (c) Convolution of the gain-bandwidth product and $|\Gamma(\omega)|^{2}$, used to infer the photodiode current noise from the measured RF power, as described in the main text.}
\label{sfig4}
\end{figure}

Fig.~\ref{sfig4}a is the measured gain as a function of frequency for
the amplifier chain and filters.  Fig.~\ref{sfig4}b shows the measured
fraction of reflected power, $|\Gamma|^{2}$, for the photodiode over
the same frequency range, obtained with the approach in
Fig.~\ref{sfig2}.  Fig.~\ref{sfig4}c shows the convolution of the
gain-bandwidth product and $|\Gamma|^{2}$.  As shown in the previous
section, this convolution may be used to infer the original mean
square current fluctuations, $i_{s}^{2}$, from the photodiode noise
source.  The results of this procedure are shown in Fig.~3 of the main
manuscript text.

\begin{figure}[h!]
\includegraphics[clip, width=8cm]{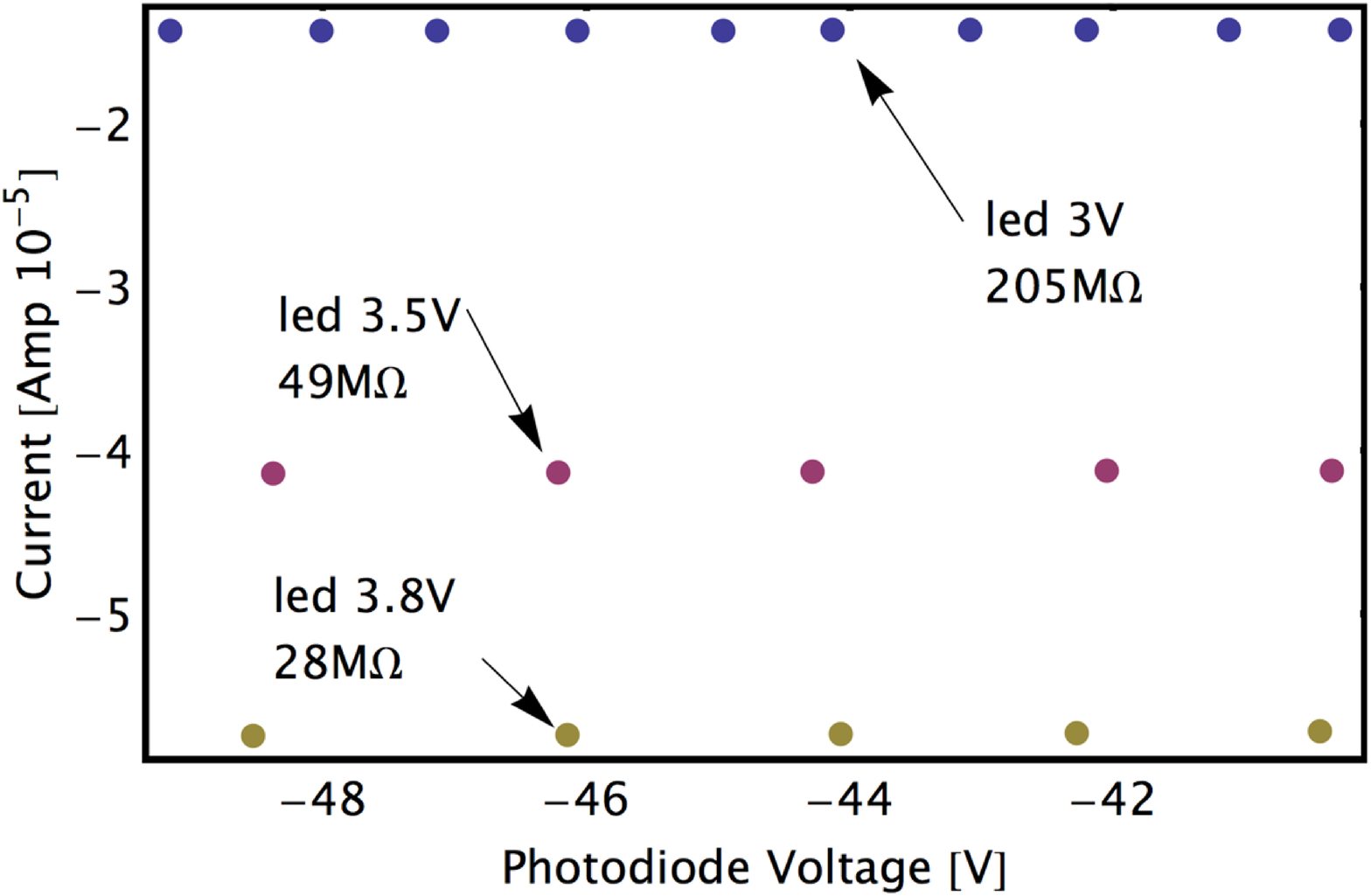}
\caption{Current-voltage characteristics of the photodiode at 
three different illumination levels (indicated by LED voltages)
near the photodiode DC operating point of -45~V.  Differential
resistances at -45~V are indicated.}
\label{sfig5}
\end{figure}

Fig.~\ref{sfig5} shows DC current-voltage characteristics of the 
photodiode around its DC operating point (-45~V) for three of the 
illumination levels (labeled by the voltages applied to the
illuminating LED) used in the photodiode shot noise measurements.
The DC resistance is always much larger than 50~$\Omega$, but as
explained in the main text, this is irrelevant to the measurements at
hand.  What does matter is the RF response of the terminated
photodiode, and as the $|\Gamma|^{2}$ data demonstrate, the 
impedance mismatch over the bandwidth of interest is not nearly as
severe as one would infer from purely DC measurements.

\section{Reflection measurements with break junctions}

\begin{figure}[h!]
\includegraphics[clip, width=12cm]{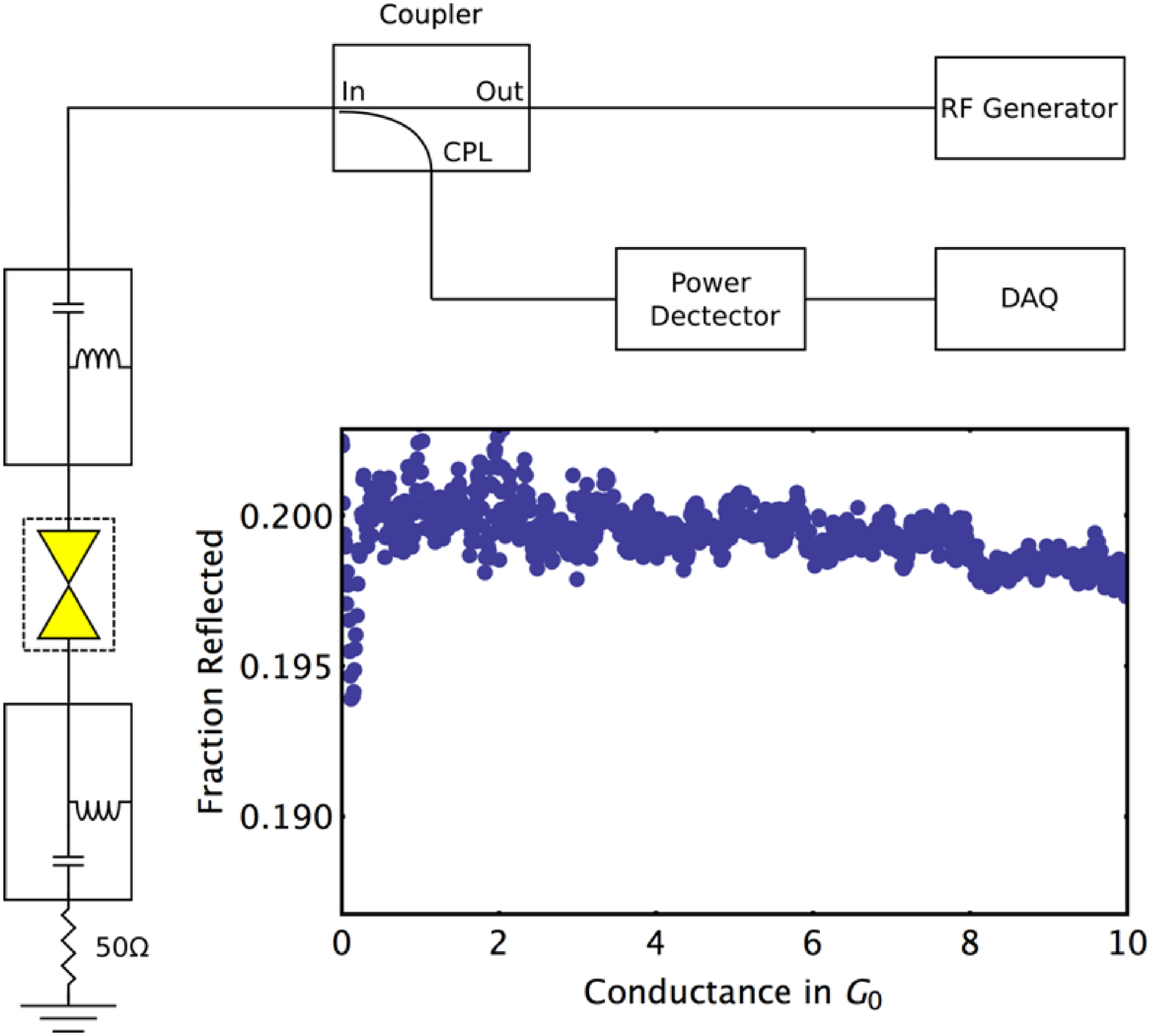}
\caption{(a) Schematic diagram for measuring the reflectance of the 
junction devices.  Not shown are the rest of the low frequency
electronics that are hooked to the DC inputs of the bias-tees, as in Fig.~1 of the main manuscript. (b) Reflectance vs. $G$ at 300~MHz.}
\label{sfig6}
\end{figure}

Figure~\ref{sfig6} shows the circuit configuration for reflectance 
measurements on the break junctions, as well as the data for
$|\Gamma|^{2}$ as a function of $G$ for a single frequency, 300~MHz.
As discussed in the main text, performing such measurements at 
many discrete frequencies across the full bandwidth is very tedius.
The main point to observe here is that $|\Gamma|^{2}$ is essentially
independent of  $G$ when $G > \sim 0.5 G_{0}$.  

\end{document}